\documentclass[a4paper]{llncs}
\usepackage[top=2.05in, bottom=2.05in, left=1.73in, right=1.73in]{geometry}
\usepackage{amssymb}
\usepackage{graphicx}
\usepackage{subfigure}
\usepackage{url}
\urldef{\mailsa}\path|{alfred.hofmann, ursula.barth, ingrid.haas, frank.holzwarth,|
\urldef{\mailsb}\path|anna.kramer, leonie.kunz, christine.reiss, nicole.sator,|
\urldef{\mailsc}\path|erika.siebert-cole, peter.strasser, lncs}@springer.com|    
\newcommand{\keywords}[1]{\par\addvspace\baselineskip
\noindent\keywordname\enspace\ignorespaces#1}

\usepackage{tabularx}
\newcolumntype{L}[1]{>{\raggedright\arraybackslash}p{#1}}
\newcolumntype{C}[1]{>{\centering\arraybackslash}p{#1}}
\newcolumntype{R}[1]{>{\raggedleft\arraybackslash}p{#1}}

\usepackage{booktabs}
\usepackage{cite}
\usepackage{subfigure}
\usepackage{amsmath}
\usepackage{graphicx}
\begin{document}

\mainmatter  % start of an individual contribution

\title{Parametric Scaling of Preprocessing assisted U-net Architecture for Improvised Retinal Vessel Segmentation}

%\titlerunning{Automated retinal vessel segmentation based on morphological preprocessing and 2D-Gabor wavelets}

\author{}
\author{Kundan Kumar$^1$\thanks{corresponding author}        \and Sumanshu Agarwal$^2$%
%\thanks{Please note that the LNCS Editorial assumes that all authors have used
%the western naming convention, with given names preceding surnames. This determines
%the structure of the names in the running heads and the author index.}%
}
%
%\authorrunning{Automated retinal vessel segmentation based on morphological preprocessing and 2D-Gabor wavelets}
%\authorrunning{}

%\institute{ECE Department, ITER, Siksha 'O' Anusandhan (Deemed to be University), Bhubaneswar, Odisha, India\\
%\email{erkundanec@gmail.com, debashishsamal@soa.ac.in},\\home page:
%\texttt{https://sites.google.com/site/erkundanec/home}}
\institute{}
\institute{Center for the Internet of Things, ITER, Siksha ‘O’ Anusandhan (Deemed to be University), Bhubaneswar-751030, Odisha, India
 \and Department of Electronics and Communication Engineering, ITER, Siksha ‘O’ Anusandhan (Deemed to be University), Bhubaneswar-751030, Odisha, India\\\email{kundankumar@soa.ac.in, sumanshuagarwal@soa.ac.in}\\
 \url{https://erkundanec.github.io/} 
}


%\toctitle{Lecture Notes in Computer Science}
\tocauthor{Authors' Instructions}
\maketitle

% such as diabetic retinopathy, glaucoma, age-related macular diseases, cardiovascular diseases, etc. 
\begin{abstract}
Extracting blood vessels from retinal fundus images plays a decisive role in diagnosing the progression in pertinent diseases. In medical image analysis, vessel extraction is a semantic binary segmentation problem, where blood vasculature needs to be extracted from the background. Here, we present an image enhancement technique based on the morphological preprocessing coupled with a scaled U-net architecture. Despite a relatively less number of trainable network parameters, the scaled version of U-net architecture provides better performance compare to other methods in the domain. We validated the proposed method on retinal fundus images from the DRIVE database. A significant improvement as compared to the other algorithms in the domain, in terms of the area under ROC curve ($>0.9762$) and classification accuracy ($>95.47\%$) are evident from the results. Furthermore, the proposed method is resistant to the central vessel reflex while sensitive to detect blood vessels in the presence of background items viz. exudates, optic disc, and fovea.
\end{abstract}

\section{Introduction}
\label{intro}
\begin{figure}[!h]
    \centering
    \includegraphics[width=0.65\textwidth]{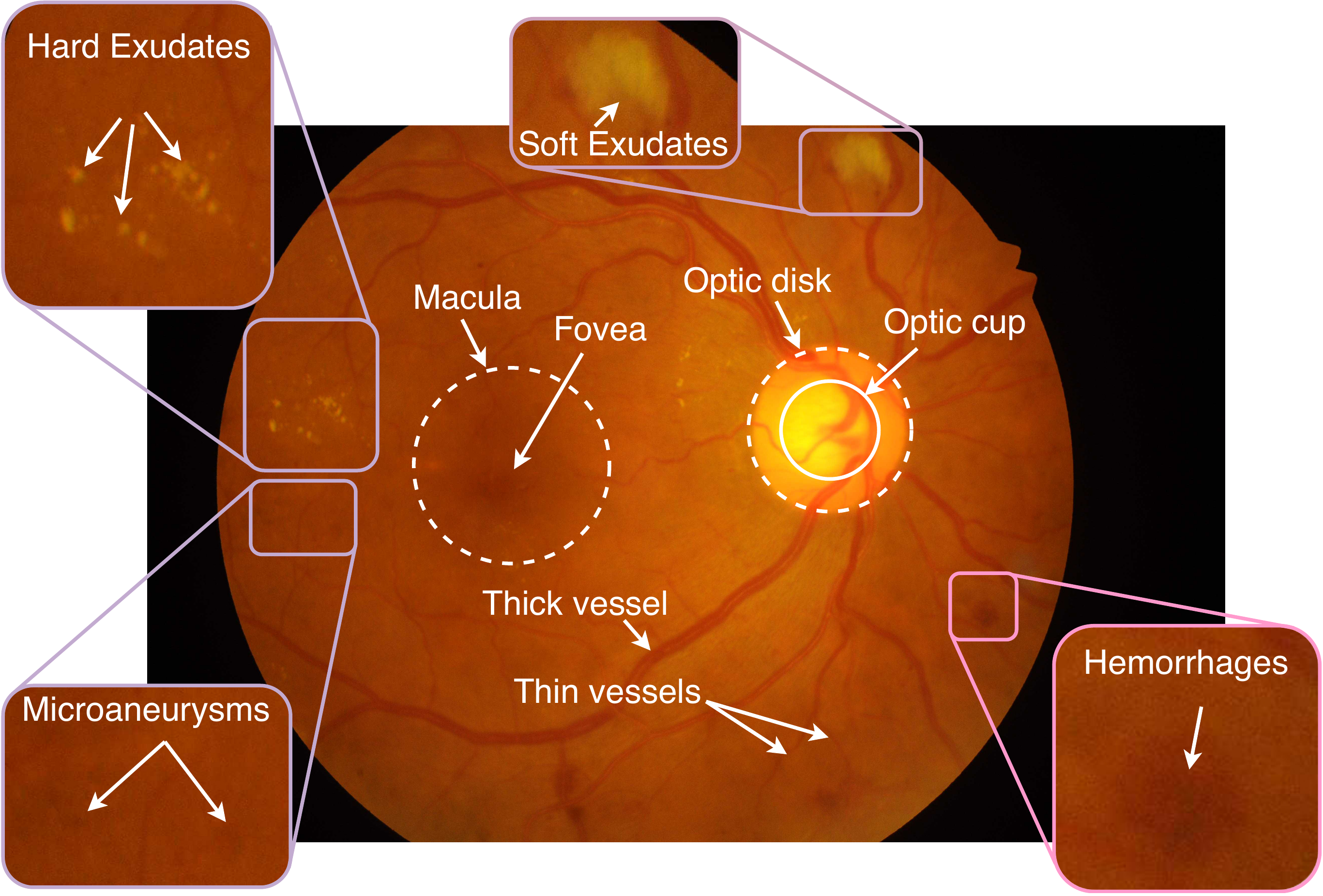}
    \caption{Labelled diagram of Retinal fundus image}
    \label{fig:labelledImg}
\end{figure}

Retinal fundus imaging is a pathological procedure used by medical professionals to diagnose various diseases. Here, the posterior portion of the human eye that contains blood vasculature map (tree-like structure) is captured using ophthalmic camera~\cite{Almotiri2018}. Retinal vasculature map helps an ophthalmologist to detect and diagnose numerous diseases, such as diabetic retinopathy~\cite{Soares2006,Hoover2000}, hypertension~\cite{Leung2004}, glaucoma~\cite{Mitchell2005}, age-related macular degeneration, which have direct or indirect relation with the brain and heart diseases~\cite{Zhang2010}. By tracking the growth and change in vascular map of retinal fundus images, timely treatment for the progression of pertinent diseases could be identified. The extraction of retinal blood vessels from the background in presence of other structures, such as exudates, optic disk, hemorrhages, microaneurysms, etc. (shown in Fig.~\ref{fig:labelledImg}), makes it more challenging and difficult.

The extraction of retinal blood vessels (RBV) can be done by an image processing tool that can handle the poor contrast between vessel and non-vessel pixels, central vessel reflex, uneven illuminance, noise, etc.~\cite{Narasimha-Iyer2008}. Many RBV extraction algorithms have been proposed in the literature to overcome these challenges. Nevertheless, an algorithm that could address all of these challenges is yet to be reported. Further, the features are manually extracted before the classification, and those features are not self-sufficient to handle all the challenges together~\cite{Fraz2012}. Conventional methods are broadly classified as (a) filtering-based methods, (b) mathematical morphology, (c) trace-based methods, (d) multiscale based approaches, (e) machine learning-based methods. Here, the first four methods are purely based on image processing tools, whereas the methods belonging to the last class, i.e. machine learning based methods, recognize the underlying patterns using man-made features dependent classification algorithms. Machine learning methods can be further sub-categorized as supervised and unsupervised learning. While supervised methods learn the classification criteria by using ground truth labelled image of retinal vessels, unsupervised methods segment the blood vessels without knowing the class information of blood vessel and background pixels. 

Among the various machine learning techniques, deep learning (DL) has proven their capability in many application such as image classification, medical image analysis, scene understanding, object detection, pose estimation, etc.~\cite{Gu2015}. The recent development in the field of deep learning algorithms are encouraging, and the technique has become very popular to solve many real-life problems. For example, convolutional deep neural network (CNN) has been endorsed in medical image processing to solve most of the segmentation problems and thereby make the system automated by eliminating the step of manual feature extraction using the conventional methods. Researchers have reviewed state-of-art automated retinal analysis based on DL~\cite{Braovic2018,Samuel2019}.
\begin{figure*}[htbp]
\centering
\includegraphics[width=0.7\textwidth]{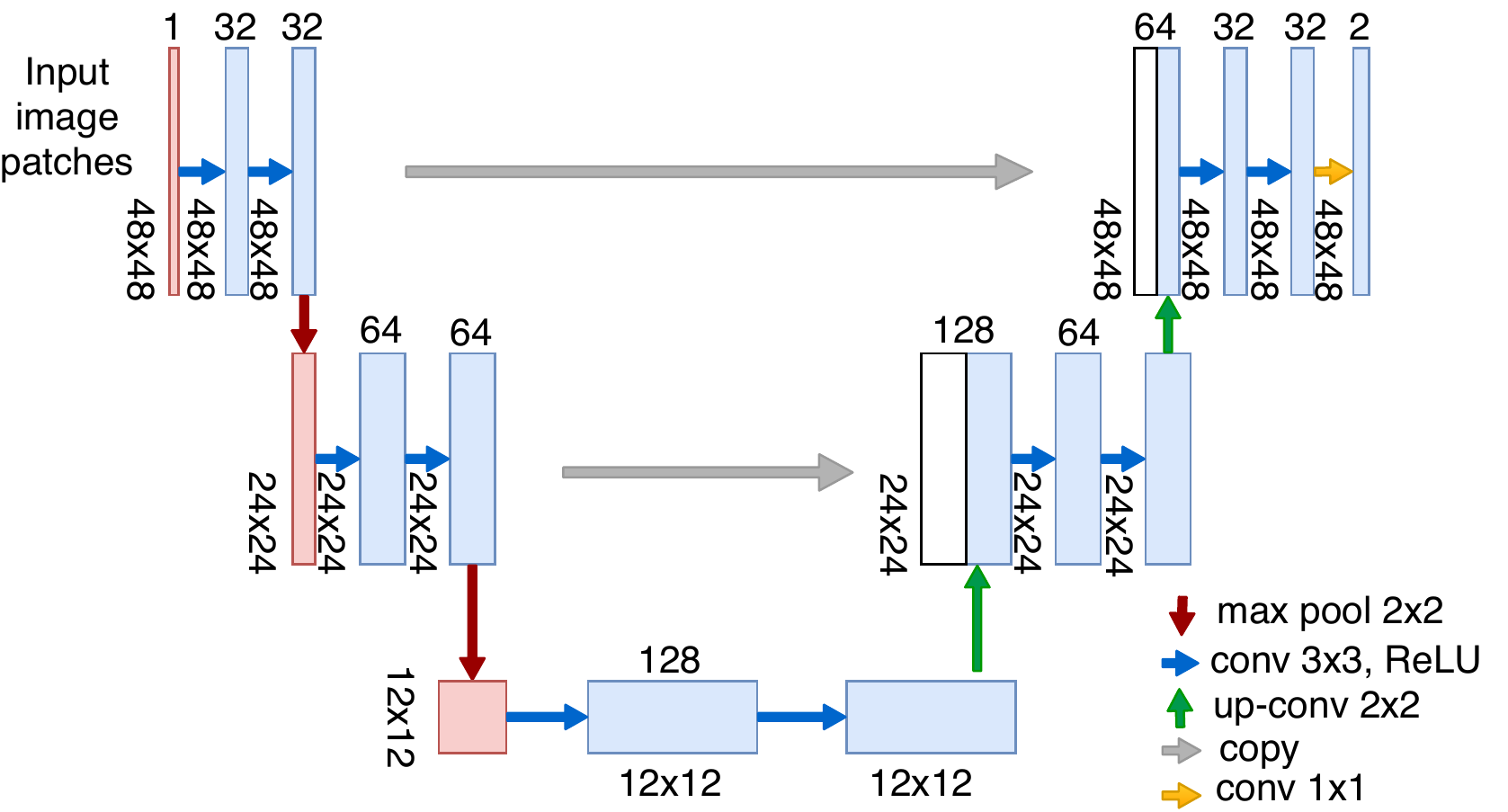}
\caption{Scaled U-net architecture}
\label{propArch}
\end{figure*}

Various convolutional neural network architectures have been introduced to handle the automated retinal analysis problem to achieve the state-of-the-art findings. The following architectures have been used to automate the vessel detection in fundus images, but not limited to, deformable U-Net (DUNet)~\cite{Jin2019a}, DeepVessel~\cite{Fu2016a}, IterNet~\cite{Li2020}.  Furthermore, fully connected convolutional neural network (FCN)~\cite{Long2015}, SegNet~\cite{Badrinarayanan2017}, U-net~\cite{Ronneberger2015}, Recurrent Residual U-net (R2U-net)~\cite{Alom2018} are well-accepted methods to segment the medical images because the architectures are based on end-to-end training. Among all of these, U-net is widely used in many applications like scene understanding, self-driving car, vegetation monitoring, medical image analysis, etc. Though U-net architecture is advantageous due to the requirement of a minimal amount of data for training, it needs excessive data augmentation to train the model.

That said, here, we present a scaled version of U-net model (see Fig.~\ref{propArch}) for RBV segmentation by using comparatively less number of learning parameters. We find that a reduction in the number of trainable network parameters leads to the complexity reduction without compromising the performance. For the study, patches extracted from the preprocessed retinal images are used. In the preprocessing step, we perform following in the order: a) grayscale conversion, b) top-hat transformation, and c) contrast limited adaptive histogram equalization (CLAHE), to enhance the contrast and uneven illuminance due to background structures like, exudates, optic disc, fovea, etc. Thereafter, a large number of image patches are randomly extracted from the preprocessed images to train the model. The major contribution of the paper can be summarized as follows:
\begin{itemize}
    \item White top-hat transformation based preprocessing is adopted to enhance the blood vessel pixel along with suppressing the other structures such as fovea, optic disk, exudates, etc.
    \item Introduced parametric scaling of U-net architecture
(scaled down U-net architecture) to reduce the number of learning parameters to fasten the learning
process.
\end{itemize}
The model suggested in the paper is validated on widely adopted retinal fundus images for segmentation from DRIVE database~\cite{Staal2004}, which is briefly discussed in the next section, where preprocessing, augmentation, and scaled version of the U-net model are also described in detail. Section~\ref{results} discusses the experimental results and the conclusions are summarized in section~\ref{conclusion}.

\section{Materials and Method}
\label{mater}
\subsection{Datasets and material}
To assess different models, the experiments were performed on DRIVE~\cite{Staal2004} dataset (a publicly available dataset). The DRIVE (Digital Retinal Images for Vessel Extraction) dataset contains 40 color images which includes 20 training and 20 test image having a plane resolution of $565\times584$. Each retinal image is an 8-bit color image. For the test set, two subsets of manually segmented ground truth images (from two different observers) are provided. The manually segmented images from the first observer are regarded as the ground truth or gold standard to assess the performance.

%\section{Method}
%\label{metho}
The segmentation is carried out in three steps: a) preprocessing and data augmentation, b) network architecture design and training, and c) postprocessing. However, in the present paper, we discuss the first two aspects of the segmentation. It should be noted that the postprocessing can further improve the performance and the work on the same is under progress.  
\subsection{Preprocessing and data augmentation}
We considered single-channel retinal images for vessel segmentation. Accordingly, the color retinal images are converted to grayscale by using the following conversion rule ($value = 0.30*R+0.59*G+0.11*B$), where $R$, $G$, and $B$ represent red, green, and blue components of the color images, respectively. Here, it is worth to mention that the green channel image can also be used instead of grayscale conversion due to its high contrast. The grayscale conversion was followed by a negative operation. Thereafter, a morphological operation (white top-hat transformation~\cite{Kumar2020}) was applied to reduce the non-uniform illumination from the image. It has been well established that morphological operations reduce the contrast of the whole image~\cite{Kumar2020}. Therefore, for image contrast enhancement, CLAHE was used.

\begin{figure}[!h]
    \centering
    \includegraphics[width = 0.33\textwidth]{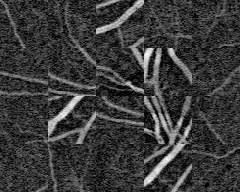}~~~~
    \includegraphics[width = 0.33\textwidth]{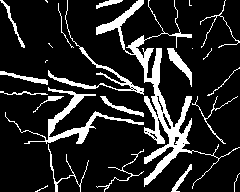}
    \caption{Few sample of preprocessed patches with its corresponding ground truth}
    \label{fig:patches}
\end{figure}

The training processes of the CNN model (discussed in the next subsection) takes a considerable amount of time for large images. Additionally, the CNN model needs plenty of images to be trained. Given that, after preprocessing, $9500$ patches of size $48\times 48$ were extracted randomly from the FOV of each image, resulting in a total collection of $9500 \times 20 = 190,000$ image patches. Few sample image patches are shown in Fig.~\ref{fig:patches}.
10\% of the total patches were used for validation, and rest image patches are used for training. Validation set is used to retain the best learned model weight during training process. For the test image, overlapped image patches of the same resolution $48\times 48$ were extracted at a stride of 5 pixels.
\begin{equation}
{\rm No.~of~test~patch} = \left( {\frac{{H - h}}{{s}} + 1} \right) \times \left( {\frac{{W - w}}{{s}} + 1} \right)
\end{equation}

Here, $H\times W$ is the retinal image resolution, $h\times w$ is the patch size, and $s$ is the stride.
For the DRIVE dataset image, $109\times 105=11445$ image patches were extracted for each test images. Note that, before extracting image patches, the test images were passed through the same preprocessing step as applied to train images.

\subsection{Architecture design and training}
In this section, we discuss the proposed CNN model as shown in Fig.~\ref{propArch}, which is a scaled version of the U-net architecture developed by Olaf Ronneberger \textit{et al.} for biomedical image segmentation~\cite{Ronneberger2015}. Similar to the full-size U-net architecture, the proposed architecture consists of two paths: encoder (named as contraction path) and decoder (known as expanding path). The encoder is very similar to the conventional stacked encoder which contracts as the network go deeper and deeper. It consists of fully connected convolutional layers and max-pooling layers for better understanding of the larger receptive field. The receptive field is a small portion of the input to produce only one node in a feature map. Before the down-sampling, response of the previous layer is processed using two convolution operation with $3\times 3$ kernels with 1 padding to the operated image. There is a rectified linear unit (ReLU) after each convolution layer. A $2\times 2$ max pooling operation with stride 2 is also used. Here, only 3 depth level architecture is modeled to design a less complex architecture with less number of parameters to be learned.  At level 3, the lowest resolution of $12\times 12$ pixels is obtained to learn the local information in the image patches. The first red box at level 1 is the input layer which takes the single-channel (grayscale) image as an input. Remaining all boxes refers to multi-scale feature maps. The size of each feature map is given on the left side of the box (see Fig.~\ref{propArch}), and the number of channel is shown on the top of the boxes. In the decoder path at each label, white boxes and up-sampled blue boxes are concatenated to regenerate the feature map at each level. To project the feature map to the number of classes, we use $1\times 1$ convolutional kernel at the last layer. After all the processing, the segmentation map is obtained at the output. It is worth to mention here that the size of image patches should be chosen in such a way that all $2\times 2$  max-pooling operations can be applied with an even x- and y-size.

% \subsection{}

In the present study, the network was trained using image patches and corresponding segmentation map (ground truth) of size $48\times 48$. We train the network from scratch starting from weight parameter initialization. The initialization of weights in a deep neural network is extremely important to avoid the excessive activation problem~\cite{Ronneberger2015}. Therefore, initial weights are drawn randomly from a Gaussian distribution, $\mathcal{N}(0,\sqrt{2/N})$, where $N$ denotes the number of incoming nodes of one neuron~\cite{He2015}. ADAM optimizer is used to update the weights in each iteration. At the beginning of training, the learning rate is initialized with 0.001 and after every 20 epochs learning rate is reduced at 0.9 decay factor.

%\subsection{Postprocessing}
\section{Experimental Results and discussion}
\label{results}
\begin{figure}[!h]
\centering
\includegraphics[width=0.45\textwidth]{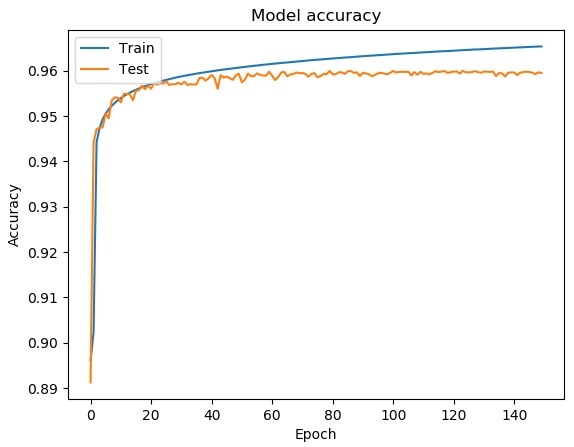}~~
\includegraphics[width=0.45\textwidth]{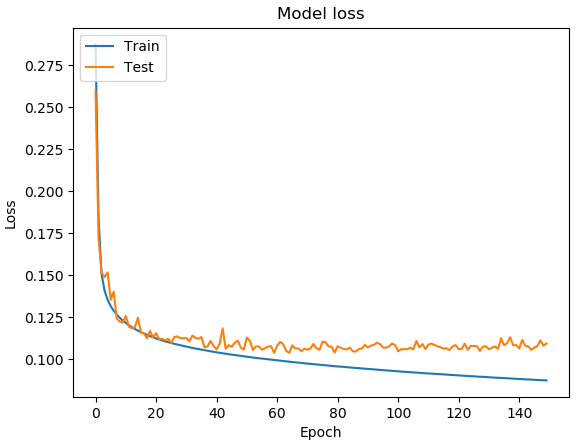}
\caption{Accuracy curve (left) and loss curve (right) of the model for training and testing}
\label{Fig_AccPlot}
\vspace{-12pt}
\end{figure}

\begin{figure}[h]
\centering
\includegraphics[scale=0.14]{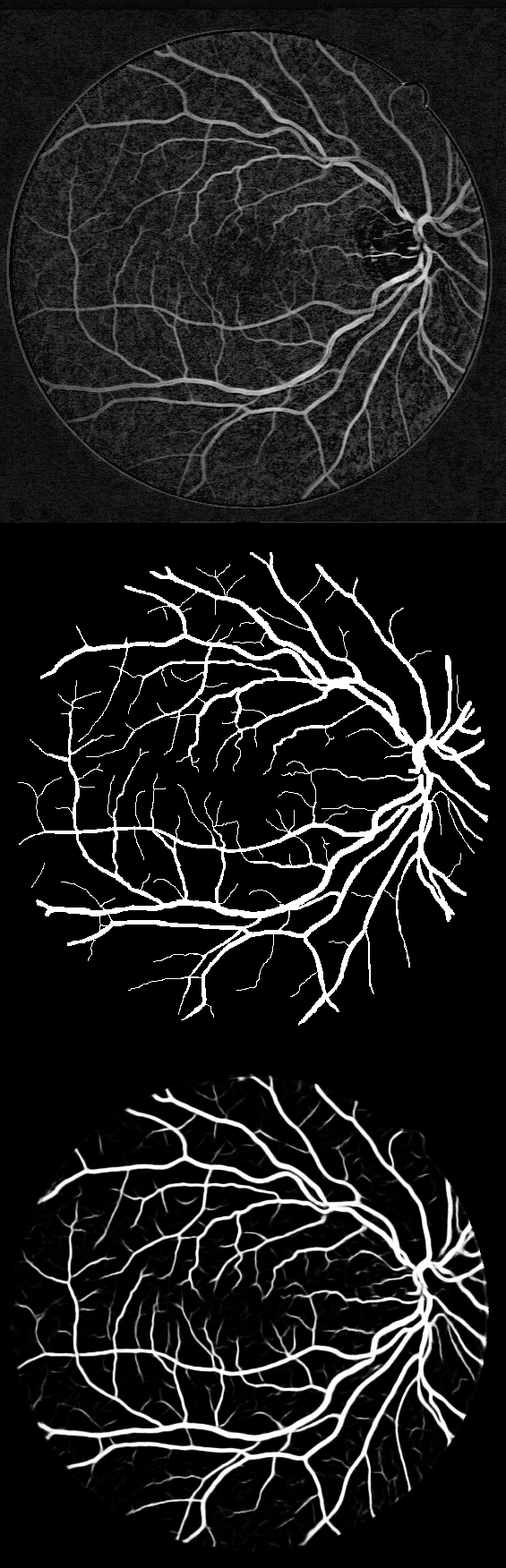}~
\includegraphics[scale=0.14]{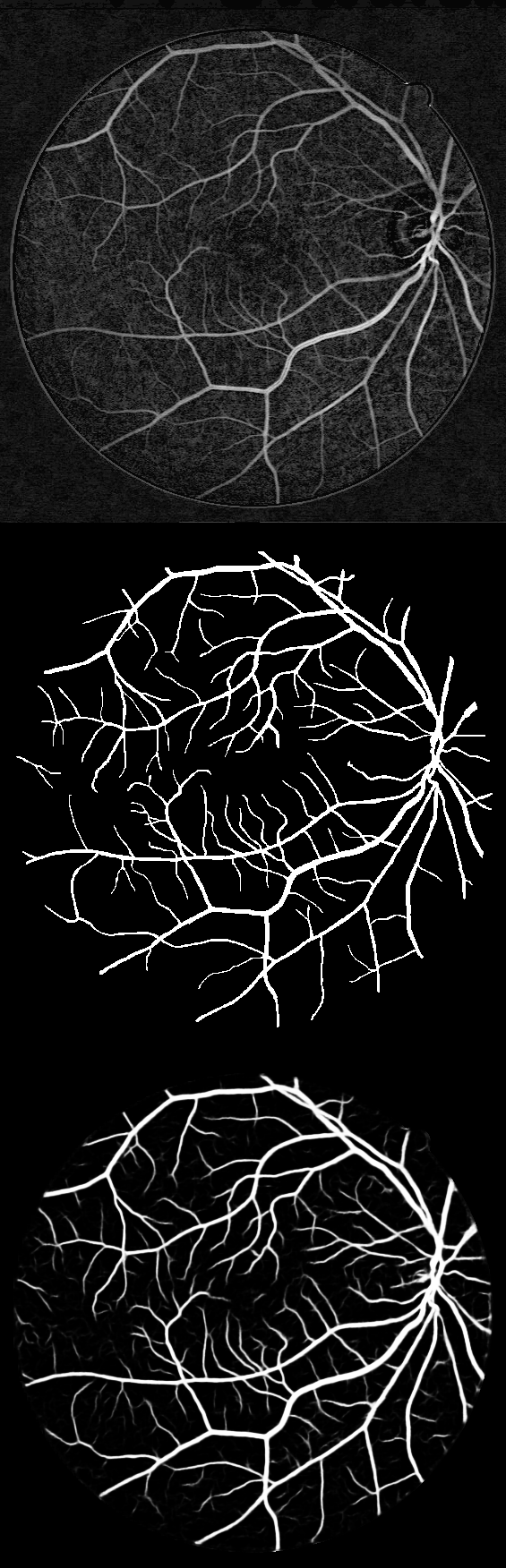}~
\includegraphics[scale=0.14]{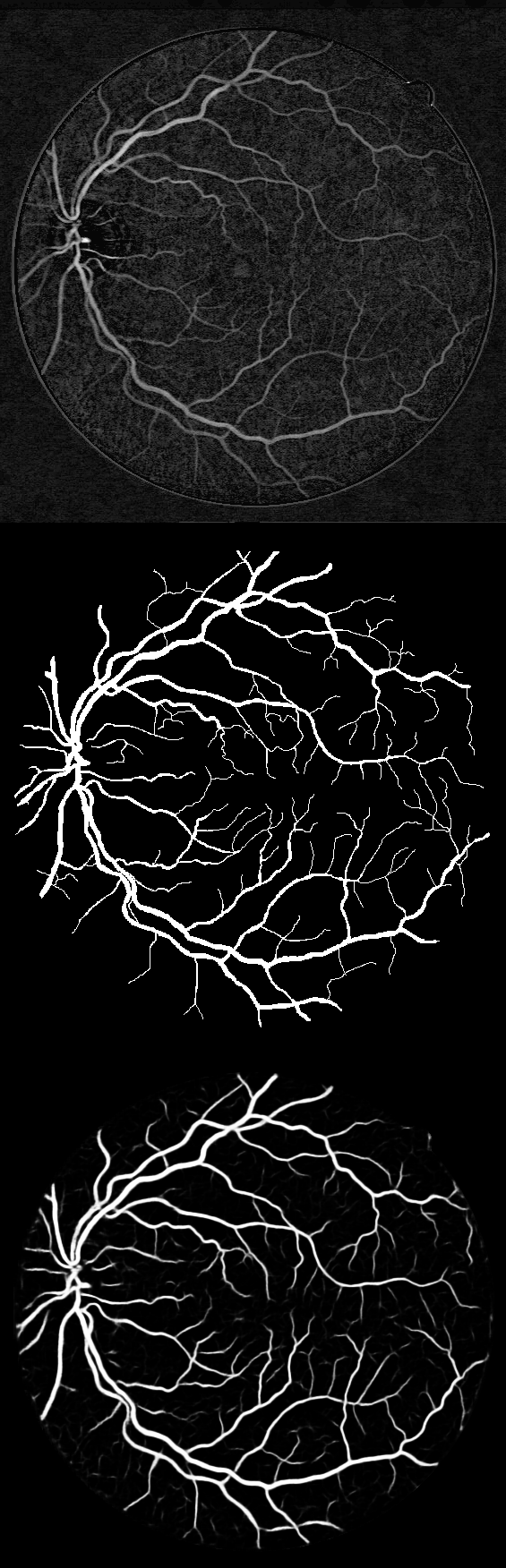}~
\includegraphics[scale=0.14]{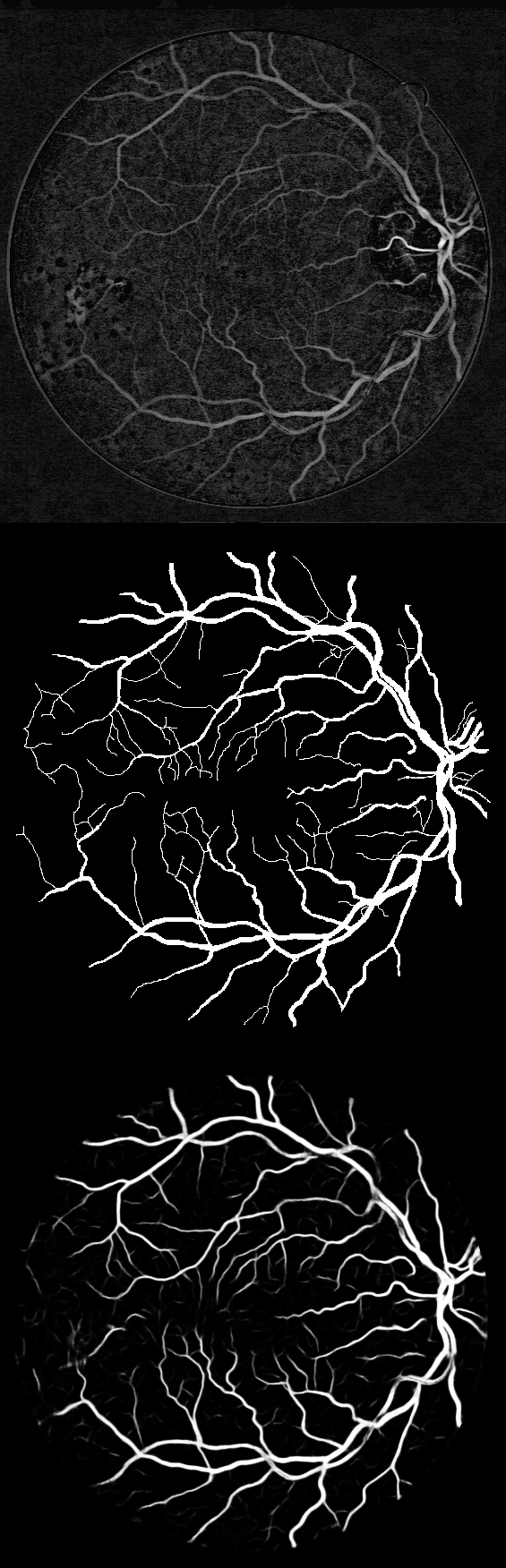}
\caption{Segmented result of test images from DRIVE dataset: (a) Preprocessed original image (first row), (b) Ground truth (second row), (c) Output response of proposed scaled U-net architecture (third row)}
\label{Fig_segOut}
\end{figure}

The proposed model was trained using extracted image patches from the training set of DRIVE dataset. The experiments were performed under Tensorflow~\cite{Abadi2016} and Keras~\cite{Chollet2015} framework on the hardware configuration  NVIDIA GeForce GTX 1080Ti GPU. In this work, the random weights were initialized in each layer. The preprocessed patches were normalized and used to train the network for 100 epochs. During training, 32 image patches were used in a single batch to compute the average error at the output, and that error is backpropagated in the network to update the weights. Accuracy and loss curves are shown in Fig.~\ref{Fig_AccPlot}. It shows that preprocessing assisted U-net architecture achieves more than 95\% accuracy in just 15 epochs.
Segmented binary maps of blood vessels and background for the retinal images taken from DRIVE dataset are shown in Fig.~\ref{Fig_segOut}.
The first row is the preprocessed retinal images, and the corresponding gold standard ground truth is shown in the second row. The final response of the proposed model is shown in the third row of the figure. The fourth column of the figure corresponds to the pathological image; however, the first three columns correspond to normal retinal images. The figure indicates that the proposed model segment the blood vessels with high accuracy, even for pathological images.
%in presence of exudates, optic disc, and fovea.
Furthermore, the proposed method is resistant to the central vessel reflex. After performing many experiments, we found that use of top-hat transformation as morphological preprocessing reduces the computation time (takes less number of iterations) to train the proposed U-net architecture for the same findings as performed by many state-of-art algorithms on the DRIVE database.

We performed the quantitative analysis in terms of  area under the curve (AUC) of receiver operating characteristic (ROC), accuracy (Acc), and sensitivity (Sen). Table~\ref{tab_quant} compares results obtained from the scaled U-net and state-of-art vessel segmentation methods evaluated on DRIVE dataset. It shows that the proposed method has better AUC, accuracy, and sensitivity compare to many algorithms in the domain.

% \begin{figure}[h]
% \centering
% \includegraphics[scale=0.45\textwidth]{Figures/ErrorPlot.png}
% \caption{Accuracy vs iteration}
% \label{Fig_AccPlot}
% \end{figure}

\begin{table}[!h]
\caption{Experimental results of proposed method for RBV segmentation and comparison with other conventional and Deep Learning based approaches}
\label{tab_quant}
\centering
\begin{tabular}{p{6cm}p{1.5cm}p{1.5cm}p{1.5cm}}
\toprule
Method        & AUC    & Acc  &  Sen \\ \midrule
2nd Observer & -      & 0.9473& - \\
Soares \textit{et al.} (2006)~\cite{Soares2006}  & 0.9614 & 0.9466 &- \\
Ricci \textit{et al.} (2007)~\cite{Ricci2007}  & 0.9558 & 0.9563 &-\\
Azzopardi \textit{et al.} (2015)~\cite{Azzopardi2015}& 0.9614 & 0.9442&0.7526\\
Liskowski \textit{et al.} (2016)~\cite{Liskowski2016} & 0.9720 & 0.9495&0.7763\\
Fu \textit{et al. } (2016)~\cite{Fu2016a}&  - & 0.9523 & 0.7603\\
Alom \textit{et. al.} (2018) (U-net)~\cite{Alom2018} & 0.9755 & 0.9531&0.7537\\
Proposed model (scaled U-net) &0.9762 &0.9547&0.7830\\\toprule
\end{tabular}
\end{table}
\section{Conclusion}
\label{conclusion}
To summarize, here, a scaled version of U-net architecture has been proposed for retinal blood vessel segmentation. Morphological processing based preprocessing has been introduced to enhance the segmentation performance and computational burden. The proposed model was evaluated using DRIVE database. The experimental results manifest that scaled version of U-net performed better segmentation with less number of learning parameters compare to existing methods applied on retinal fundus images; however, morphological preprocessing reduces the run-time for the network training process. Furthermore, the proposed method is resistant to the central vessel reflex while sensitive to detect blood vessels even if background structures viz. exudates, optic disc, fovea, etc. are present.

\bibliographystyle{splncs03_unsrt}
\bibliography{mybib.bib}
\end{document}